\documentclass{article}

\usepackage{arxiv}

\usepackage[utf8]{inputenc} 
\usepackage[T1]{fontenc}    
\usepackage{hyperref}       
\usepackage{url}            
\usepackage{booktabs}       
\usepackage{amsfonts}       
\usepackage{nicefrac}       
\usepackage{microtype}      
\usepackage{lipsum}		
\usepackage{graphicx}
\usepackage{doi}
\usepackage{caption}
\usepackage{subcaption}

\title{Classification of anomalous gait using Machine Learning techniques and embedded sensors}

\author{{Tiago R.~de~Sá} \\
	Electrical Engineering\\
	University of State of Amazonas\\
	Manaus, AM\\
	\texttt{trs.eng17@uea.edu.br} \\
	\And
	{~Carlos M. S.~Figueiredo} \\
	PhD, Computer Science\\
	University of State of Amazonas\\
	Manaus, AM\\
	\texttt{cfigueiredo@uea.edu.br} \\
}

\renewcommand{\shorttitle}{\textit{arXiv} Template}

\hypersetup{
pdftitle={Classification of anomalous gait using Machine Learning techniques and embedded sensors},
pdfsubject={machine learning},
pdfauthor={Tiago R.~de~Sá, ~Carlos M. S.~Figueiredo},
pdfkeywords={Human Gait, Machine Learning, Internet Of Things, Wearables, Healthcare}
}

\begin{document}
\maketitle

\begin{abstract}
Human gait can be a predictive factor for detecting pathologies that affect human locomotion according to studies. In addition, it is known that a high investment is demanded in order to raise a traditional clinical infrastructure able to provide human gait examinations, making them unaffordable for economically vulnerable patients. In face of this scenario, this work proposes an accessible and modern solution composed of a wearable device, to acquire 3D-accelerometer and 3D-gyroscope measurements, and machine learning techniques to classify between distinct categories of induced gait disorders. In order to develop the proposed research, it was created a dataset with the target label being 4 distinct and balanced categories of anomalous gait. The machine learning techniques that achieved the best performances (in terms of accuracy) in this dataset were through the application of Principal Component Analysis algorithm following of a Support Vector Machines classifier (94 \%). Further, an architecture based on a Feedforward Neural Network yielded even better results (96 \%). Finally, it is also presented computational performance comparison between the models implemented.
\end{abstract}


\section{Introduction}
{G}ait analysis is the systematic study of human locomotion \cite{Turner:2019, Tao:2012}. Thereby, from the point of view of clinical observation, it is impossible to analyze and quantify its degree of deviation from normality without adequate instruments. So, gait analysis, using clinical instruments available in a motion analysis laboratory, provides relevant data for the treatment of pathologies that affect human locomotion \cite{Chen:2012}. Consequently a high investment is requested in a traditional clinical infrastructure. In parallel to this context, with the continuous advancement of the scientific community regarding Internet of Things and modern techniques of Digital Signal Processing based on Machine Learning approaches, excellent progress has been seen in applications in the healthcare domain \cite{Majumder:2017, Ukil:2016, Daniela:2016, slijepcevic2020explanation}, especially with regard to the analysis of the gait of individuals based on data provided by low-cost embedded sensors such as accelerometer and gyroscope \cite{Potluri:2019, Bosio:2017, Catalfamo:2010, Takeda:2009}.

The implantation of a motion analysis laboratory demands a high investment with state-of-the-art technologies for a complete analysis of the parameters that characterize human gait, making clinical exams less affordable. In view of the above, according to studies \cite{Davarzani:2020, Hannink:2017, Majumder:2017, Muro:2014, Daniel:2010}, the main components needed in a traditional motion analysis laboratory are: a video system for capturing movement through markers \cite{Saucier:2019, Chen:2012}; equipment system for electromyographic analysis; equipment system for static and dynamic capture of plantar pressure data; software for analyzing three-dimensional motion and, finally, physical space. Thereby, the investment to raise such infrastructure is not less than a few tens of millions of dollars \cite{Silva:2011}. In this context, more economically viable alternatives have been developed to capture the data needed to analyze human gait \cite{Tao:2012}.

Among the alternatives above, the use of wearable sensors such as accelerometer \cite{masiala2019featuresetengineering} and gyroscope stands out \cite{Nemes:2020}, as they allow a three-dimensional kinematic analysis of human movement \cite{Brognara:2019}. In this scenario, briefly, the process for identifying irregularities in a patient's gait is conditioned to the processing of signals data obtained from wearable sensors. Given the processed data, a possible approach is the application of machine learning techniques for feature extraction \cite{El_Maachi_2020}. Then, with a trained model, it is possible to make inferences about the parameters of a patient's gait \cite{Pardoel:2019}. Finally, according to studies \cite{Bosio:2017, Burdack_2020}, among the most common methods for movements patterns classification , stands out Support Vector Machine, Random Forest Classifiers, Multi-Layer Perceptrons and Convolutional Neural Networks.

To exemplify one of the possible clinical scenarios where the analysis of human gait parameters is relevant, it is considered the case of patients with diabetes mellitus (DM). Recent studies reveal that such patients tend to have irregularities in their gait due to a chronic complication known as diabetic peripheral neuropathy from DM \cite{MARTINELLI:2014, Saura:2010, Fregonesi:2010}. Thus, early detection of this anomaly would prevent future complications in the diabetic patient's condition, such as the reduction in the risk of falls and the consequent reduction in the appearance of ulcers.

Considering the aspects mentioned above, this work seeks (i) to contribute to the creation of a human gait dataset through the implementation of a low-cost application composed of hardware and software, and (ii) to present comparative assessments between different machine learning models applied to this dataset.

In view of these proposed objectives, section 2 presents the related works. Section 3 describes about each component of the proposed application. Section 4, in turn, presents the results obtained as well as performance comparisons for each proposed technique. Finally, section 5 presents the final considerations and suggestions for future work.

\section{Related Works}
The number of scientific contributions in understanding the parameters that characterize human gait through wearable embedded sensors and Machine Learning techniques has been increasing. Following is presented some related scientific contributions in this scenario.

In \cite{Potluri-2:2019},it was developed a data collection system composed of an embedded wearable hardware and a software implemented in Java. Then, plantar pressure data of the reference foot and kinematic data were collected to measure the orientations of the reference leg segment. Regarding the system software, it was used to add additional information such as age, weight and height to the database.

Given the work above, to estimate the parameters that characterized the patient's gait, it was implemented a model based on a stacked Long-Term Memory Network (LSTM). Through this model, it was possible to demonstrate how the behavior of a normal gait could be detected. For this detection, the characteristics of error predictions in normal gaits compared with pathological simulated gaits were considered.

In \cite{Turner:2019}, on the other hand, they described how non-invasive wearable sensors could be used in combination with Deep Learning techniques to classify artificially induced gait changes. For this purpose, the plantar pressure of the foot was measured in 12 participants, who each underwent 8 artificially induced gait changes by modifying the lower part of the shoe. Data was recorded at 100 Hz on 2520 data channels. After data collection, these were used to train a model also based on Deep Learning techniques through a stacked Long-Term Memory Network (LSTM). 

Many of the related works describe advantages and disadvantages between important materials and methods for building a database. For example, in \cite{Chen:2012}, they presented a comparative view of current quantitative measurement instruments for gait analysis. Thus, it was possible to observe the advantages of using wearable embedded sensors for the following reasons: low computational cost, low financial cost, practicality of use, satisfactory accuracy and precision. In all these points, the choice for the analysis system based on wearable sensors stood out compared to the conventional system as they require an infrastructure that is costly from a financial and computational point of view.

The proposed work, unlike the related works presented, stands out for the following reasons:

\begin{itemize}
  \item Implementation of a low-cost wearable device to collect kinematic data;
  \item Creation of a balanced gait dataset based on only two measurement units: 3D-accelerometer and 3D-gyroscope;
  \item Presentation of performance comparisons among different machine learning and deep learning approaches. 
\end{itemize}

Next, the methods and materials used in this work will be described.

\section{Material and Methods}

\begin{figure*}[!ht]
\centering
\includegraphics[width=6.5in]{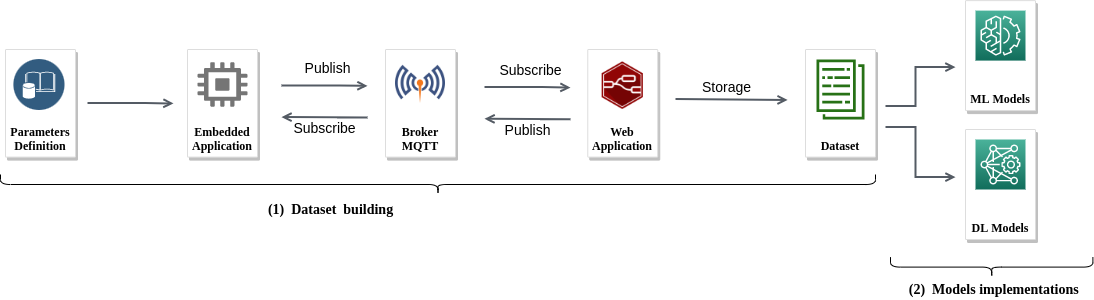}%
\caption{Components presents in the workflow of the proposed methodology}
\label{fig:Arq_Geral}
\end{figure*}

The Fig.~\ref{fig:Arq_Geral} below provides an overview of all components that will be part of the development of this work: (1) creation of the dataset and (2) implementation and comparison between machine learning models.

\subsection{Gait Dataset}
The steps to create the dataset were: (1) definition of collection parameters given the human gait context, (2) implementation of the embedded application and (3) implementation of the web application. The specifications and responsibilities of each step above are described in detail below.

\subsubsection{Dataset Description}
For proof of concept of the proposed work, 840 instances of human gait data were collected among 20 volunteers. In a balanced way, data were distributed between 4 distinct categories.

The 4 distinct categories of human gait were considered based on four different artificial modes of walking, namely, (i) walking, (ii) marching, (iii) limping and (iv) feint movements. These 4 categories were used as dependent variables to be predicted, namely, target labels.

In order to illustrate the use of the categories above when collecting data from the volunteer gait, it is asked to him to walk by simulating each of the category above. Thus, considering the first category, walking (i), the participant is instructed to walk at his common pace. Yet, considering the second category, march (ii), he is instructed to imitate, as closely as possible, the military march. The same procedure is followed for the other two remaining categories.

In all categories defined above, the following simulation criteria were pre-established: a flat surface, collection period equal to 5 (five) seconds and the reference joint being the right leg, in the ankle region. Furthermore, the data content consisted of two vector physical quantities - the three-dimensional linear acceleration and the three-dimensional angular rotation.

All participants who voluntarily contributed to the construction of the dataset were healthy, without any pathology related to locomotion.

\subsubsection{Embbeded Application}
\begin{figure}[!ht]
\centering
\includegraphics[width=2.5in]{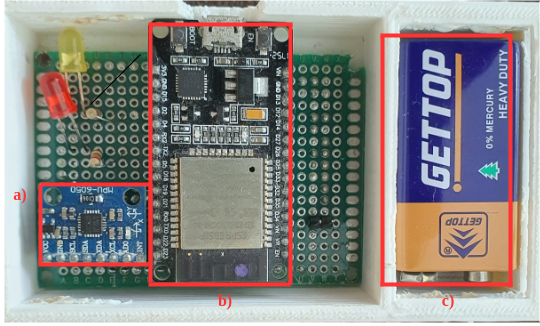}%
\caption{Prototype of the embedded wearable device. Figure a) represents the moviment sensor MPU6050, b) illustrates the microcontroller ESP32 DevKit C and c) represents the power supply.}
\label{fig:case_3d.png}
\end{figure}

\begin{figure}[!ht]
\centering
\includegraphics[width=1.5in]{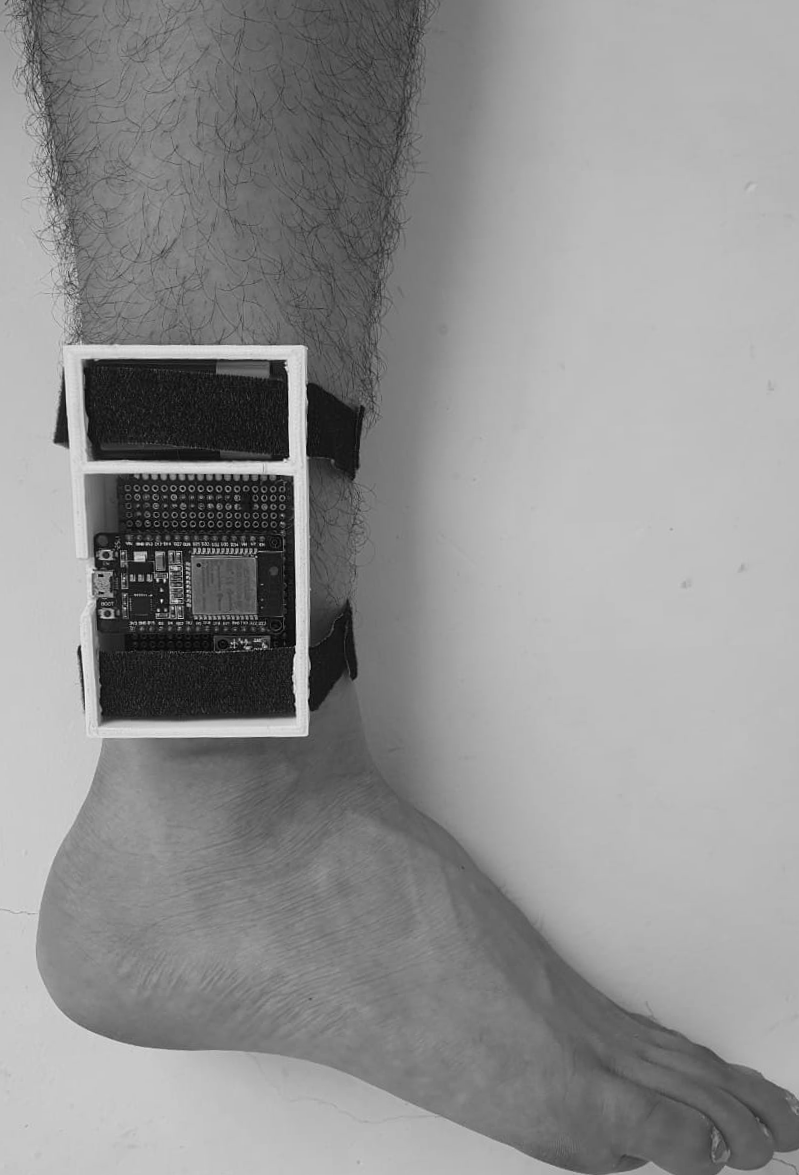}%
\caption{Prototype of the embedded wearable application fixed with velcro tape close to the ankle region.}
\label{fig:leg.png}
\end{figure}

The Fig.~\ref{fig:case_3d.png} illustrates the prototype of the proposed embedded device for data collection. This consists of a microcontroller (Fig.~\ref{fig:case_3d.png}-b), a sensor (Fig.~\ref{fig:case_3d.png}-a) and a 3D support case to fit the electrical circuit and use it in the region near the ankle (Fig.~\ref{fig:leg.png}). 

The choice of the microcontroller was made based on a survey of requirements given the context of the proposed work. This should contain, at a minimum, UART and I2C communication protocols, wireless network connectivity, embedded memory, low-power processors, in addition to input/output pins.

Thus, with regard to the UART (Universal Asynchronous Receiver/Transmitter) protocol, this would be responsible for enabling serial communication aiming at loading the embedded software developed on a host plataform to the target platform, in addition to enabling the debug of the embedded program. Still, the I2C protocol would be responsible for initializing and reading the motion sensor data through a master and slave relationship. At the same time, wireless network connectivity would make it possible to send data collected from an embedded device to an external server. Embedded memory, in turn, would provide the space to store the application code and others. Finally, the input/output pins in a minimum quantity sufficient for the peripheral connection of the kinematic sensor.

In order to meet the specifications listed above, it was opted for the ESP32 DevKit C development board as its system-on-a-chip (SoC) meets all the requirements defined above. 

The motion sensor chosen was the MPU-6050 module. This enabled the collection of three-dimensional data on physical quantities such as linear acceleration and angular rotation. Then, to enable communication between the sensor and the microcontroller, the i2cdevlib library, implemented by Jeff Rowberg, was used. Through this library, the sensor sampling frequency was defined at 100 Hz. From this definition, it was possible to collect 500 samples in each of the three-dimensional axes. That is, for each axis of linear acceleration and angular rotation, 500 samples were simultaneously collected, providing a total of 3000 (6 x 500) samples per file. In summary, these samples represented the behavior of human gait in different categories in terms of linear acceleration and angular rotation discretized in time and amplitude.

Finally, 3D case support was developed through TinkerCad software. The support was designed in a mold so that it had dimensions compatible with the designed hardware, in addition to having the characteristic of making the device wearable on the ankle, as shown in Figs.~\ref{fig:case_3d.png} and ~\ref{fig:leg.png}.

As will be seen below, after the data is collected by the embedded device, it is published in a  broker server  through mqtt protocol in order to automate the dataset labeling process.

\subsubsection{Web Application}
The proposed web application had two main purposes: labeling the data and then storing them in csv (comma-separated-values) format on the server's hard disk. The following figure illustrates one of the interfaces developed for the web application through the Node-RED framework. 

\begin{figure}[!ht]
\centering
\includegraphics[width=2.0in]{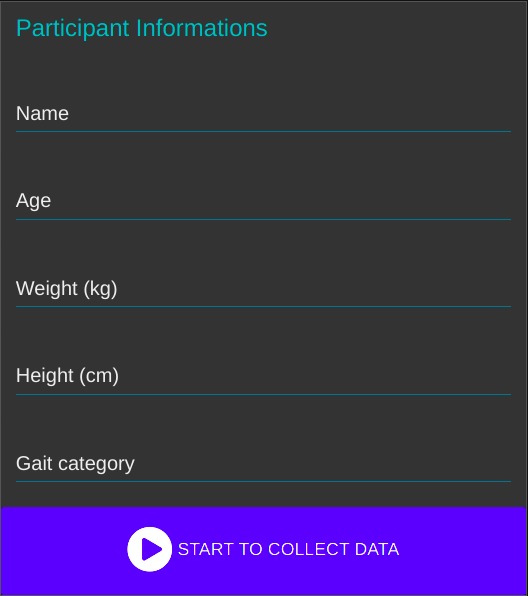}%
\caption{Implemented interface for inserting data in the supervised dataset.}
\label{fig:main_screen.png}
\end{figure}

Regarding the frontend, two interfaces were implemented. The first interface, as illustrated in Fig,~\ref{fig:main_screen.png} , allows the user to enter the following information: name, age, height, weight and the gait category that will be artificially reproduced afterwards. For the second interface, a connection test between the web application and the embedded device was implemented before starting the induced gait reproduction process.

Regarding the backend, on the other hand, the information entered by the user was used to label the file that was generated after the end of the simulation. After the labeling step, the file is properly saved according to the reproduced gait category.

Finally, the communication between the embedded application and the web application was established through the MQTT (Message Queuing Telemetry Transport) protocol and the Eclipse Mosquitto broker on a Linux server.

\subsection{Machine Learning Models}
\begin{figure}[!ht]
\centering
\includegraphics[width=3.0in]{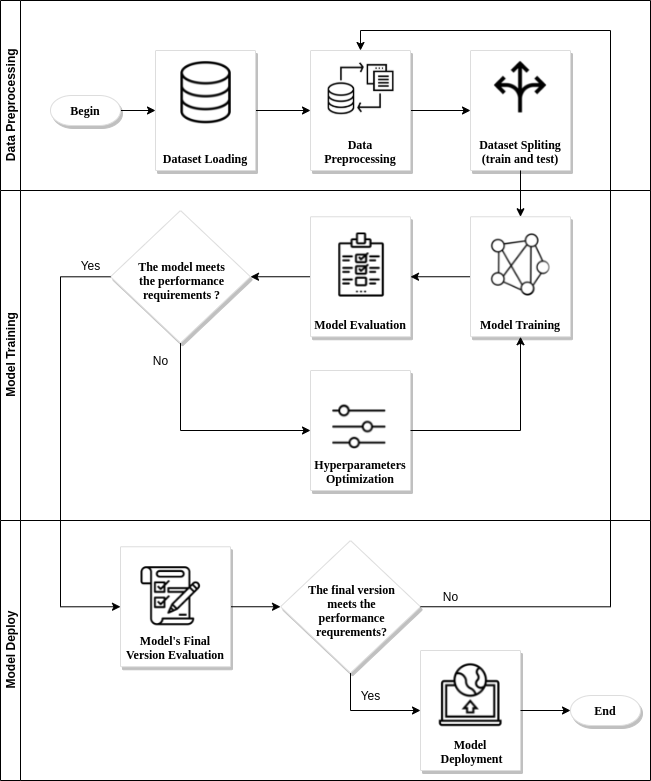}%
\caption{Defined steps for implementing the models}
\label{fig:arquitetura_ia}
\end{figure}

A total of 4 models were implemented aiming to provide insights about performance evaluation considering the dataset created. Thereby, the goal of each model is the same: classify a given instance in one of the 4 established classes. In all cases, the models were trained based on the workflow illustrated by Fig.~\ref{fig:arquitetura_ia} with the split being 20\% to the test dataset and the remaining data (80 \%) to the train dataset.

In summary, 2 models were trained with different Machine Learning algorithms - namely, Support Vector Machine (SVM) and Random Forests (RF). In addition, a model based on a Feedforward Neural Network (FNN) is presented as well as a deep learning model based on a Convolutional Neural Network (CNN),  totaling 4 models. The table~\ref{tab:Table1} provides an overview of the parameters for each chosen model.

\begin{table*}[!ht]
\caption{Summary of training configurations used for models based on Machine Learning and Deep Learning techniques.}
\centering
\includegraphics[width=6.5in]{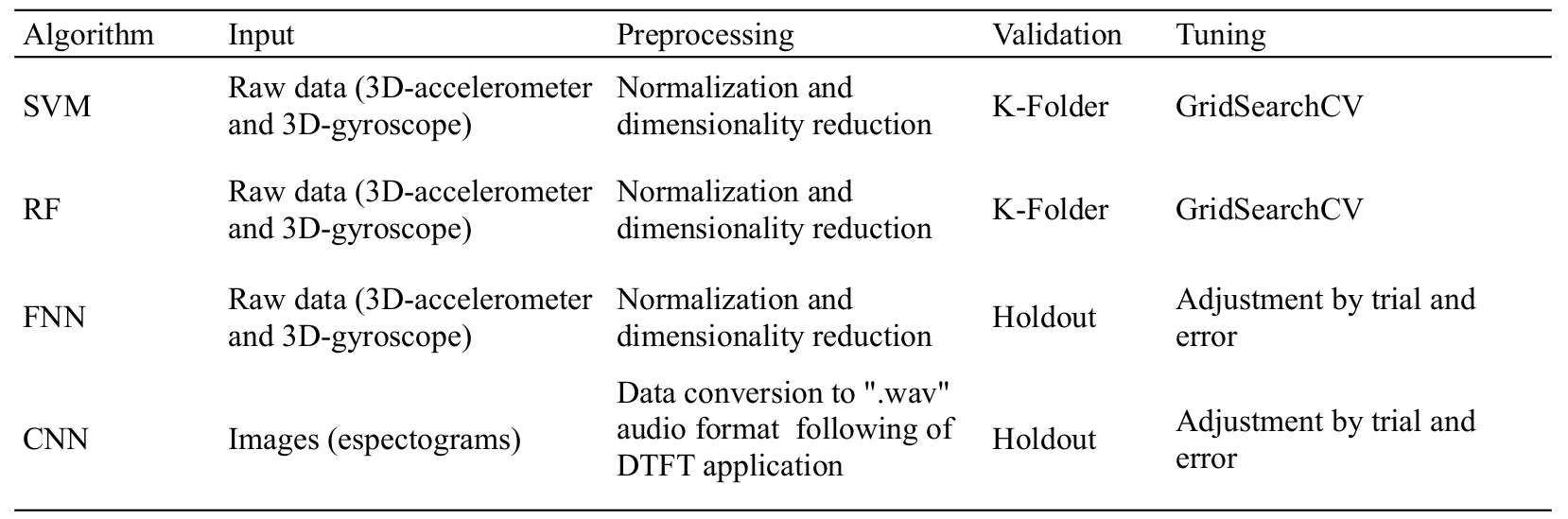}%

\label{tab:Table1}
\end{table*}

The hyperparameters (number of hidden layers, perceptrons and activation functions) of the model trained based on a FNN architecture were adjusted by trial and error. The table~\ref{tab:Table1} summarizes the main definitions and procedures performed.

The following hyperparameters were chosen for the FNN architecture:

\begin{itemize}
  \item Input Layer: 60 neurons with ReLU function activation;
  \item Hidden Layer 1: 2048 neurons with ReLU function activation;
  \item Hidden Layer 2: 1024 neurons with ReLU function activaiton;
  \item Output Layer: 4 neurons with softmax function.
\end{itemize}

\begin{figure}[!h]
\centering
\includegraphics[width=3.5in]{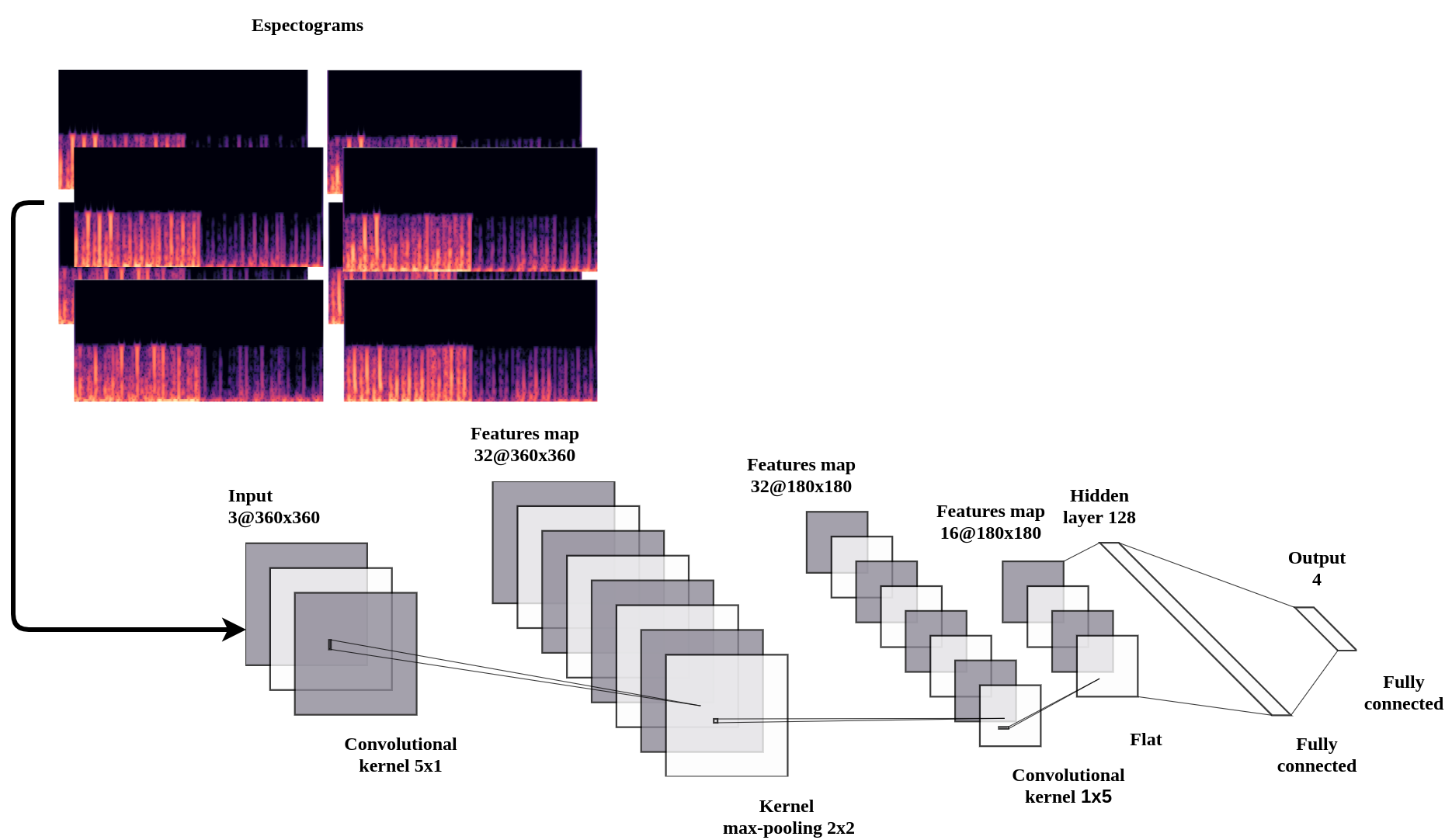}%
\caption{Neural Convolutional Network implemented using spectograms.}
\label{fig:cnn_cnn}
\end{figure}

As is illustrated by table~\ref{tab:Table1}, the model trained based on a CNN algorithm had a input considerably different. It performed predictions based on images of spectograms. They represent a snapshot of 3D-acceleration and 3D-rotation data acquired over the period considered. Fig.~\ref{fig:cnn_cnn} illustrates the CNN hyperparameters and spectograms been used as input to the model.

In order to evaluate this model, it was created a dataset containing spectograms instead of raw data. Then, to generate them, the following preprocessing steps were executed:

\begin{itemize}
  \item Concatenation of 3D-acceleration and 3D-rotation raw data for each dataset instance;
  \item Creation of audios files .wav (main lib used to this: librosa) from the concatenation above;
  \item Creation of spectograms by applying discrete-time Fourier transform (DTFT) on each audio file created;
\end{itemize}

Finally, a total of 840 spectrograms were generated, balanced between the 4 distinct categories defined previously.

Next section presents the results obtained from the models defined above.

\section{Results}
After building the dataset, efforts were focused on evaluating the models performance. The metrics chosen were accuracy, precision, recall and F1 score. In addition, it were also assessed the computational performance in terms of time required to perform a predict task for each model. The results are presented below.

\subsection{Model 1 - Support Vector Machine (SVM)}

Among the classic machine learning techniques used, the model based on SVM algorithm yielded the best performance in terms of the evaluation metrics considered.  The following hyperparameters were used after applying grid search method::

\begin{itemize}
  \item Decision function shape: One vs one;
  \item Regularization parameter: 100;
  \item Kernel coefficient: 0.01;
  \item Kernel type: radio basis function.
\end{itemize}

\begin{figure*}[!ht]
   \centering
   
 \subfloat[Model 1 learning curve.]{\label{fig:estimador_svm_learning_curve}\includegraphics[width=.4\textwidth]{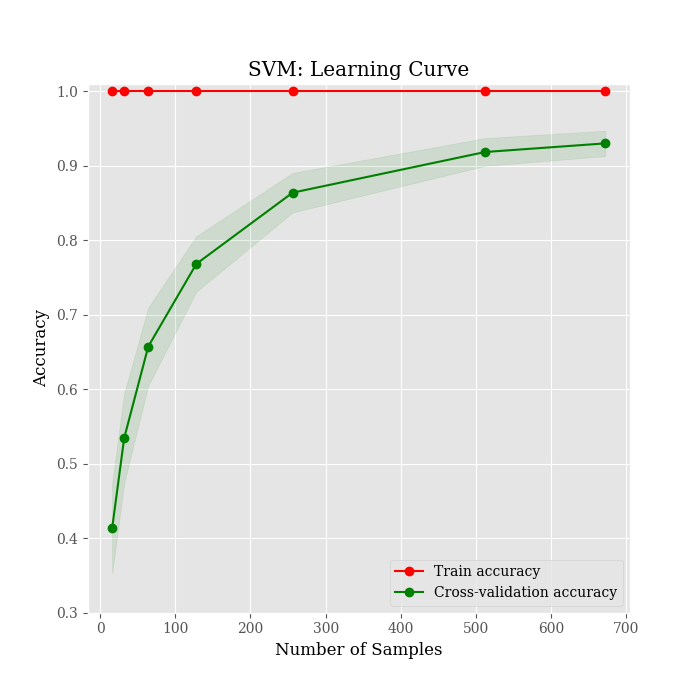}} 
 \subfloat[Model 1 confusion matrix.]{\label{fig:estimador_svm_confusion_matrix}\includegraphics[width=.4\textwidth]{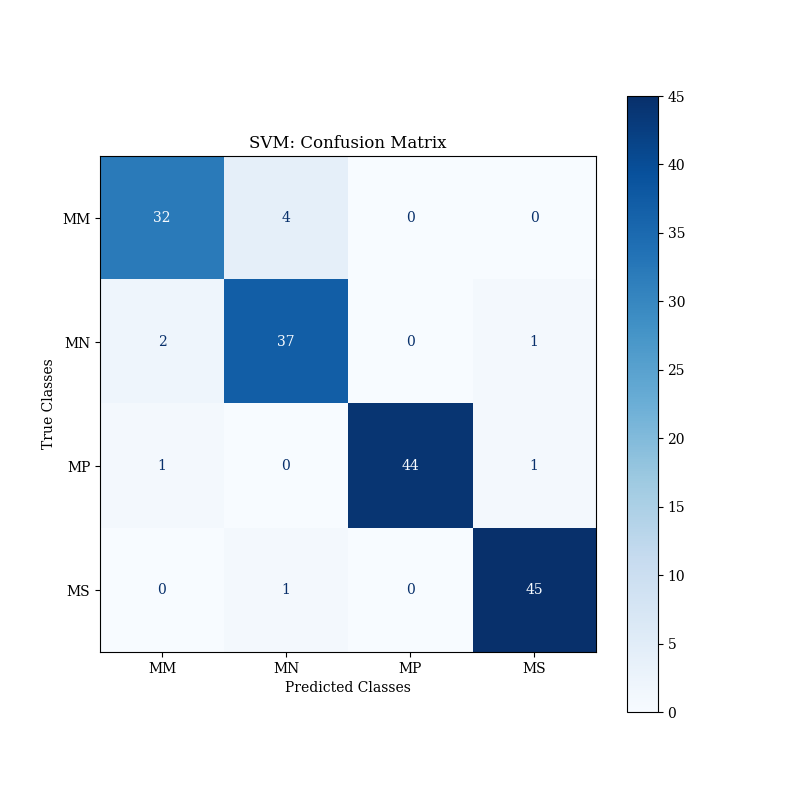}}
   \caption{Training results for the model based on the SVM classifier.}\label{fig:_estimador_svm_learning_curve}
   \end{figure*}

Fig.~\ref{fig:_estimador_svm_learning_curve} illustrates the learning curve for training the model based on the SVM classifier and presents the confusion matrix results.

The confusion matrix presented in the Fig~\ref{fig:estimador_svm_confusion_matrix} shows that some gaits reproduced in category MM (limping gait) are missclassified as MN (normal gait) and vice versa. This observation is coherent from the point of view of the empirical analysis of the reproduced movement, since, in few situations, during the dataset creation process, the difference between a normal gait and a limp gait was not evident.

Finally, for the learning curve of the model based on the SVM algorithm (Fig.~\ref{fig:estimador_svm_learning_curve}), it is possible to infer that the increasing in the number of samples (horizontal axis) does not tend to cause significant increases in accuracy metric (vertical axis). In this way, the cross-validation accuracy is numerically close to its maximum reachable  value.

\subsection{Model 2 - Random Forest (RF)}

The model trained based on Random Forest technique had the advantage of using ensemble methods. In that way, results are yielded by averaging multiple decision tree predictions and, consequently, the variance is reduced.  The following hyperparameters were used after applying grid search method:

\begin{itemize}
  \item Number of trees in the forest: 100;
  \item Criteria to measure the quality of a split: Gini;
  \item Minimum number of samples requered to split an internal node: 2;
  \item The minimum number of samples required to be at a leaf node: 1.
\end{itemize}

\begin{figure*}[!ht]
   \centering
   
 \subfloat[Model 2 learning curve.]{\label{fig:estimador_rfc_learning_curve}\includegraphics[width=.4\textwidth]{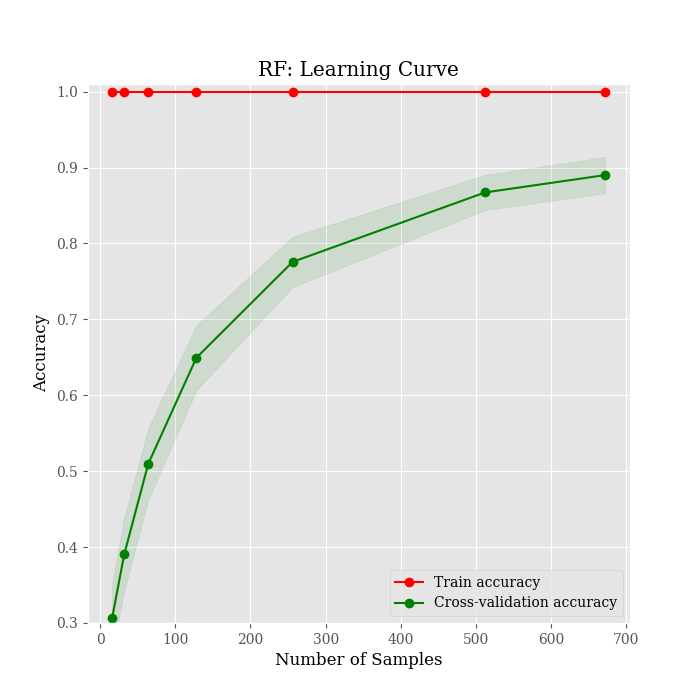}} 
 \subfloat[Model 2 confusion matrix.]{\label{fig:estimador_rfc_confusion_matrix}\includegraphics[width=.4\textwidth]{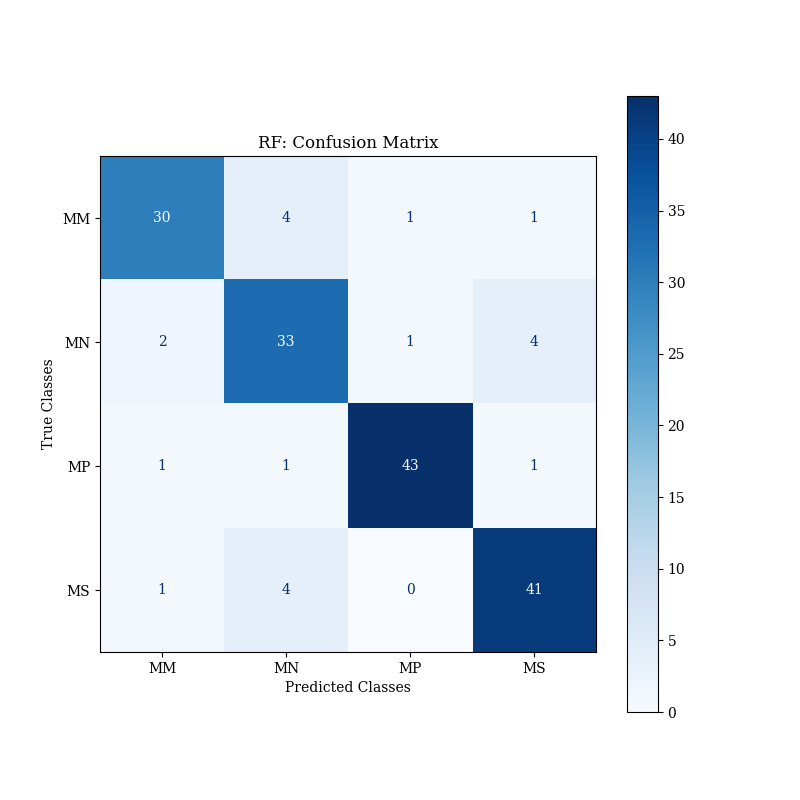}}
   \caption{Training results for the model based on the RF classifier.}\label{_fig:estimador_rfc_learning_curve}
   \end{figure*}

Fig.~\ref{_fig:estimador_rfc_learning_curve} illustrates the learning curve for training the model based on the RF classifier and presents the confusion matrix results.

The confusion matrix presented in the Fig~\ref{fig:estimador_rfc_confusion_matrix} shows that some gaits reproduced in category MM (limping gait) are also missclassified as MN (normal gait) and vice versa. In addition, missclassifications are produced between MN and MS (military gait).

Finally, considering the Fig.~\ref{fig:estimador_rfc_learning_curve}, it is possible to infer that the increasing in the number of samples could yield even better and significant results.

\subsection{Model 3 - Feedforward Neural Network (FNN)}

\begin{figure*}[!ht]
   \centering
   
 \subfloat[Model 3 learning curve.]{\label{fig:estimador_fnn_learning_curve}\includegraphics[width=.4\textwidth]{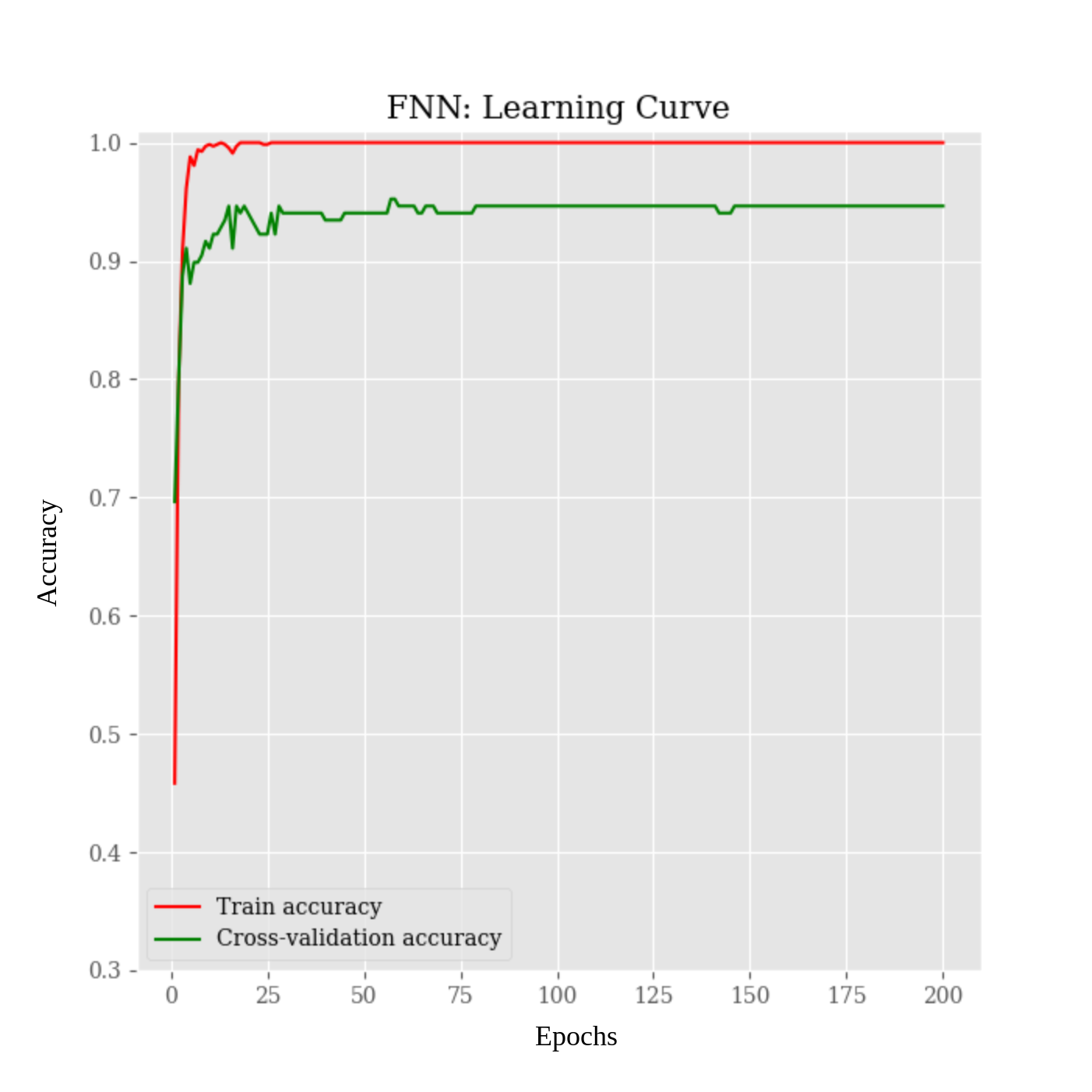}} 
 \subfloat[Model 3 confusion matrix.]{\label{fig:estimador_fnn_confusion_matrix}\includegraphics[width=.4\textwidth]{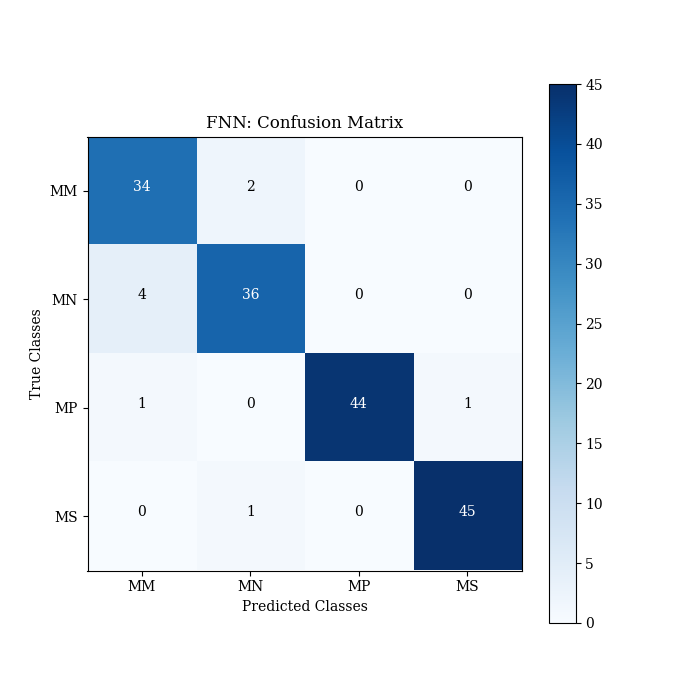}}
   \caption{Training results for the model based on FNN architecture.}\label{fig:_estimador_fnn}
   \end{figure*}

So far, models have all been trained with classic Machine Learning techniques. For model 3, however, an architecture based on a Feedforward Neural Network was used (architecture described in section 3.2). Thus, for the definition of the architecture, several tests and manual enhancements were made in the hyperparameters until the bests results were reached. Next, Fig.~\ref{fig:_estimador_fnn} illustrates the learning curve and confusion matrix of model 3.

The learning curve of model 3 reveals that the increasing in the number of training periods does not tend to cause relevant changes in the validation accuracy. However, increasing the number of samples can make the learning curve more consistent and less oscillatory.

In summary, in all the confusion matrices presented, it is observed that the category "MM" (limping gait) had the highest number of instances missclassified.

\subsection{Model 4 - Convolutional Neural Network (CNN)}

\begin{figure*}[!ht]
   \centering
   
 \subfloat[Model 4 learning curve.]{\label{fig:estimador_cnn_learning_curve}\includegraphics[width=.4\textwidth]{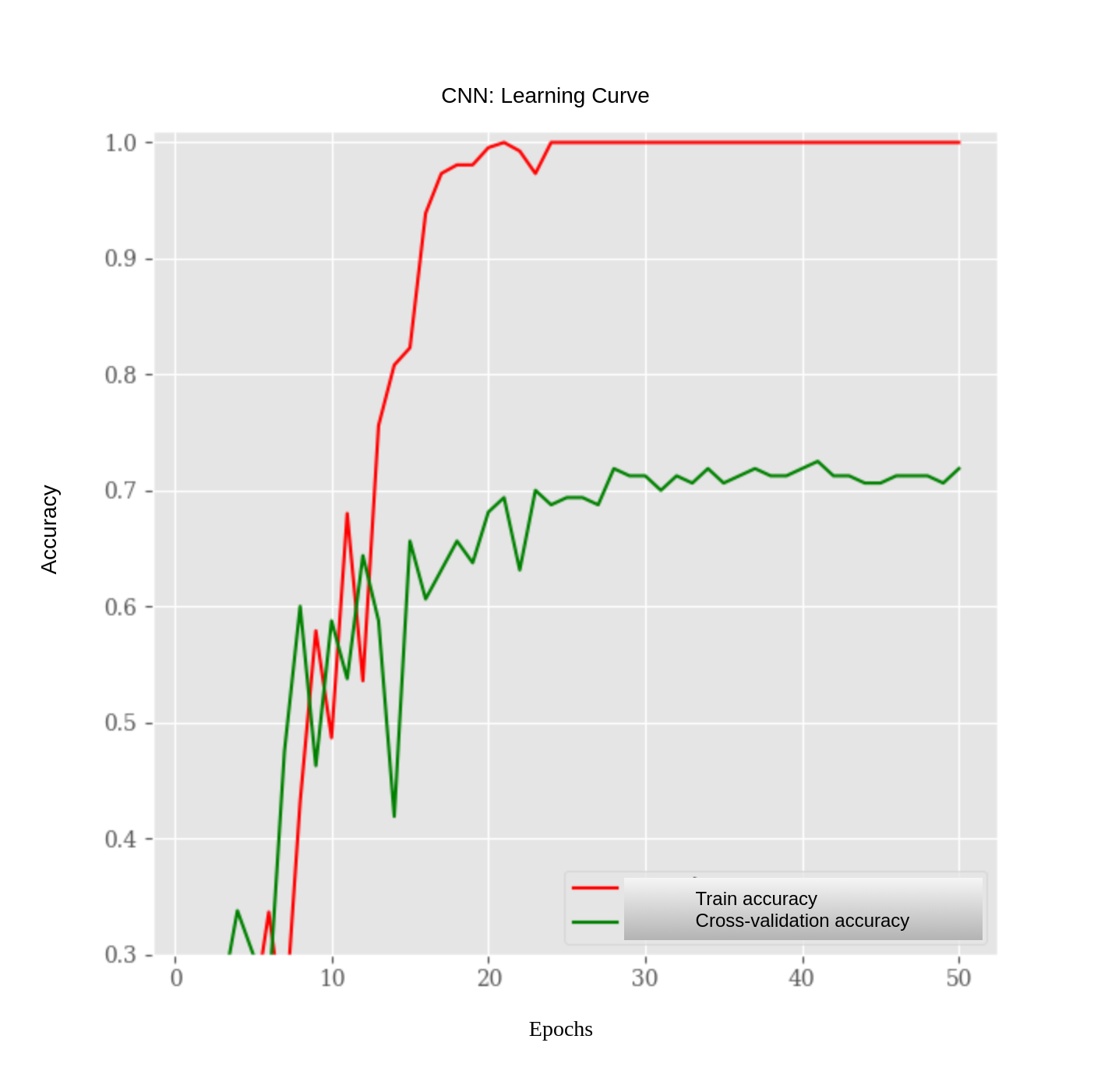}} 
 \subfloat[Model 4 confusion matrix.]{\label{fig:estimador_cnn_confusion_matrix}\includegraphics[width=.4\textwidth]{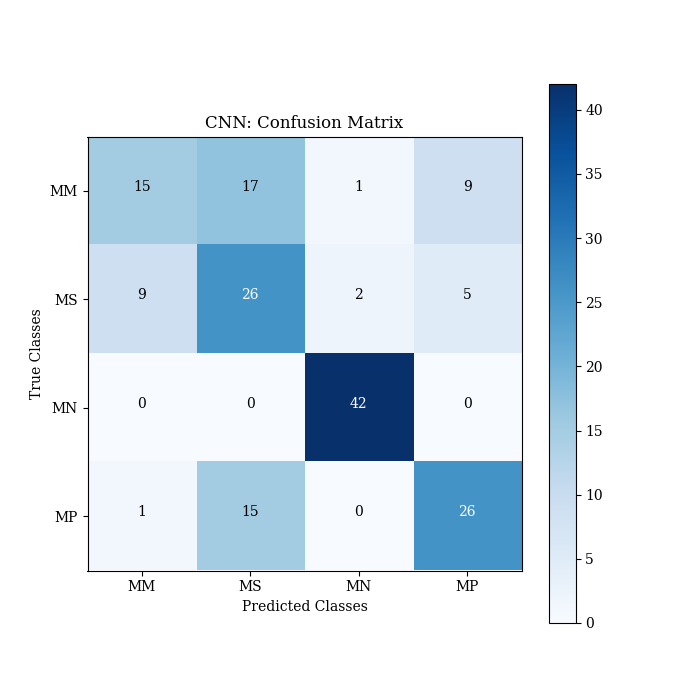}}
   \caption{Training results for the model based on CNN architecture.}\label{fig:_estimador_cnn}
   \end{figure*}

The learning curve of the trained model according to the architecture illustrated in Fig.~\ref{fig:cnn_cnn} for spectrogram classification is shown in Fig.~\ref{fig:_estimador_cnn}.

The results achieved with this approach were not satisfactory compared to the other applied techniques. Despite this, future studies may be developed in order to reach a conclusive result on the use of Convolutional Neural Networks to classify spectrograms from human gait data.

Finally, the table~\ref{tab:results} presents the results generated from each of the models trained. The results are evaluated in terms of 4 following metrics: F1 score, precision, recall and accuracy

\begin{table}[!ht]
\caption{Summary of the evaluation metrics obtained from the models above.}
	\label{tab:results}
	\centering
		\begin{tabular}{lllll} 
		\hline
	
     	&	Precision   &	Recall	& F1 Score	&	Accuracy	\\
	\hline
FNN 	&	0.94	   &	0.94	&	0.93	        &	0.96	\\
SVM 	&	0.94	   &	0.93	&	0.93	        &	0.94	\\
RF 	&	0.90	   &	0.90	&	0.90	        &	0.90	\\
CNN & 0.70 & 0.72 & 0.69 & 0.77 \\
		\hline
	
	\end{tabular}

\end{table}

The results summarized in the table above shows that the model that achieved the best performance, in terms of accuracy, was the model 3.

\subsection{Computational performance comparison between the models}

The comparative evaluation of the models in terms of prediction time was carried out with the test set (168 instances). In this way, the mean time (in milliseconds) and standard deviation required by the prediction task on the test set was performed. This procedure was executed for each model 20 times. Finally, this evaluation was performed on a machine with the following configurations:  
\begin{itemize}
  \item GPU: NVIDIA GeForce RTX 2060
  \item RAM: 16 GB
  \item SSD: 512 GB
  \item Processor: Intel Core i7 10th 2.6 GHz
\end{itemize}

\begin{table}[!ht]
\caption{Computational performance comparison between the models 1, 2, 3 and 4 in terms of milliseconds with standard deviation.}
	\label{tab:FO}
	\centering
		\begin{tabular}{lllll} 
		\hline
	
     Models	& Prediction time over the test set in milliseconds
     	\\
	\hline
Model 1 (SVM) 	&	4.18 ± 0.35 
        \\
Model 2 (RF) 	&	8.05 ± 0.72	 
        \\
Model 3 (FNN) 	&	28.48 ± 14.60 	
        \\
Model 4 (CNN) & 702.73 ± 12.19 	
        \\
		\hline
	
	\end{tabular}

\end{table}

\section{Conclusion}

The application presented in this work contributed to the implementation of a low-cost wearable device to collect kinematic data from human gait. The data were used in the creation of a balanced human gait dataset in 4 distinct categories, composed by only two measurement units: 3D-acceleration and 3D-rotation. Then, as seen, it was presented performance comparisons among different machine learning and deep learning approaches. The models that achieved the best results were model 1 (SVM - 94 \%) and model 2 (FNN - 96 \%). Furthermore, although the second model had yielded better results than model 1, its computational cost, in terms of prediction time, is much higher than this model. Finally, through these satisfactory results, future research can be carried out using the proposed work as a starting point.


\end{document}